\documentstyle[12pt, a4,fleqn]{article}
\begin{document}
\hyphenation{anti-fermion}
\topmargin= -20mm
\textheight= 230mm
\baselineskip = 0.33 in
 \begin{center}
\begin{large}
 {\bf {  Hyperfine Structure Constants for Eu Isotopes : 
 
 Is The Empirical Formula of HFS Anomaly Universal ? }}

\end{large}

\vspace{2cm}

T. ASAGA\footnote{e-mail: tasaga@shotgun.phys.cst.nihon-u.ac.jp}, 
T. FUJITA\footnote{e-mail: fffujita@phys.cst.nihon-u.ac.jp} and 
K. ITO\footnote{e-mail: kito@shotgun.phys.cst.nihon-u.ac.jp}

Department of Physics, Faculty of Science and Technology  
  
Nihon University, Tokyo, Japan

\vspace{3cm}

{\large ABSTRACT} 

\end{center}

We calculate the  hyperfine structure constant for the Eu isotopes 
with  shell model wave functions. The calculated 
results are compared with those predicted by 
the Moskowitz-Lombardi (M-L) empirical formula. It turns out that the 
two  approaches give the very different behaviors 
of the hfs constants in the 
isotope dependence. This should be easily measured by experiment, 
which may lead to the universality check of the M-L formula. 

\vspace{1cm}

\newpage

\begin{enumerate}
\item{\Large Introduction}

The magnetic hyperfine structure constant (hfs)  has been extensively 
studied for various nuclei since it can present some 
interesting information on the magnetization distribution in nucleus [1-5]. 
The main difference between the magnetic moment operator and 
the magnetic hfs constant lies in the appearance of the 
new type of operator $ \Sigma_i^{(1)}$ as defined by 
$$ \Sigma_i^{(1)} \equiv s_i + \sqrt{2\pi} [sY^{(2)}]_i^{(1)}   .
\eqno{(1.1)} $$
This operator $ \Sigma_i^{(1)}$ looks similar to the magnetic 
moment operator. However, it presents often quite different 
behaviors from the magnetic moment. In particular, this  
shows up in the isotope shifts of the hfs constant. For example, 
the isotope shifts of the hfs constant for the Mercury nuclei 
indicate that the expectation value of the $ \Sigma_i^{(1)}$ operator 
is constant over wide range of the shell model configurations. 

Due to this fact, there is a remarkably good empirical formula 
which is proposed by Moskowitz and Lombardi (M-L formula) [5]. 
This M-L rule simply states that the isotope shifts of the 
hfs constant for the Hg nuclei can be well described if we write the 
hfs anomaly $\epsilon$ as 
$$ \epsilon = {\alpha \over{\mu}} \eqno{(1.2)} $$ 
where $\alpha = \pm 0.01 $ n.m. for the valence neutrons 
with the spin of  $I=\ell \pm {1\over 2}$, and 
$\mu$ denotes the magnetic moment. The M-L formula can describe the 
isotope shifts $\Delta_{12}$ of the hfs constant 
for the Hg nuclei surprisingly well. 

Now, a question arises. Can the M-L formula  work equally well 
for other nuclear isotopes as well ?  This is the main question 
we want to address in this paper. 

Recently, Werth et al.[6] 
proposed to measure the isotope shifts $\Delta_{12}$ 
of the hfs constant for 
the Eu nuclei. This is a very interesting isotope from the point of view 
of the empirical formula. The measured magnetic moment of 
the Eu nuclei shows a drastic change at the mass number $A=153$. 
For the isotopes of the Eu from $A=145$ to 151, the magnetic moment 
may well be described by the core polarization effects. 
However, the magnetic moments of $A=153$ and the heavier Eu nuclei 
 are roughly one half of those 
of the lighter Eu isotopes.  

This big change of the magnetic moments should show up if we 
use the empirical formula of eq.(1.1).  Indeed, if we calculate 
the isotope shifts $\Delta_{12}$ 
of the hfs constant for the Eu nuclei using the M-L 
empirical formula of eq.(1.2), then we obtain 
a sharp transition at $A=153$ for the $\Delta_{12}$. 

On the other hand, we can also calculate the hfs constant 
of the Eu isotopes by using the shell model wave functions. 
This is what we have done in this paper. We calculate the magnetic 
hfs constant of the Eu nuclei by considering the core  polarization 
effects. For the big change of the magnetic moment at $A=153$, 
we consider some special state which is assumed to absorb the 
magnetic moment. This state must be connected to some collective 
state. But here we do not argue in detail which kind of properties 
this special state can possess, apart from the assumption 
that this state has a vanishing magnetic moment.  Under this assumption, 
we obtain the isotope shifts  $\Delta_{12}$ for the Eu nuclei 
which do not show any sudden change at $A=153$. 
This is mainly because the core polarization effects on the 
magnetic moment and on the $\Sigma_i^{(1)}$ operators behave very 
similarly  and thus they cancel with each other. This large 
difference between the two theories should be checked by experiment. 

From the experimental side, there is an important  
 progress to measure 
the isotope shifts of the magnetic hfs constant [6,7].  This becomes 
possible due to the ion trap method which can isolate the atoms. 
This ion trap method can measure  hfs separations  with an accuracy 
of $10^{-8}$ or even lower, which is by far better than  the accuracy any 
theoretical models can predict. Also, the feasibility to measure 
directly nuclear magnetic moments by the ion trapping technique 
has been demonstrated successfully[8]. 
With this high accuracy of the ion trap method, there is 
some possibility to check the time reversal invariance 
in the atomic processes once nuclear ambiguities are removed. 
This is also one of our purpose to study the magnetic hfs constant,
 though we still do not have any good physical quantities at hand to 
 study the time reversal invariance.

The paper is organized as follows. In the next section, we briefly 
explain the theory of the magnetic hyperfine structure 
in electronic atoms. 
Then, section 3 treats the calculation of the nuclear part of the hfs 
constant by using shell model wave functions. Also, we explain a special state 
which has  a vanishing magnetic moment. 
 In section 4,  numerical results of the isotope 
 shifts of the hfs constant for the Eu nuclei are presented.  
 Also, the empirical formula of Moskowitz and Lombardi is 
 compared to the present calculation. In section 5, we 
 summarize what we have understood  from 
this work. 

\vspace{3cm}

\item{\Large Magnetic Hyperfine Structure Constant}

The atomic electron which is bound by the nucleus feels 
 the magnetic interaction in addition to the static Coulomb force. 
The magnetic interaction between the electron and the nucleus 
 can be described as
$$ H' = -\int {\bf j}_N ({\bf r}) {\bf A}({\bf r}) d^3 r  \eqno{(2.1)} $$
where the nuclear current ${\bf j}_N ({\bf r})$ can be written as 
$$ {\bf j}_N({\bf r}) = {e\hbar\over{2Mc}}\sum_i g_s^{(i)}
\nabla \times s_i \delta ({\bf r}-{\bf R}_i ) 
+\sum_i g_{\ell}^{(i)} {e\over{2M}}( {\bf P}_i \delta ({\bf r}-{\bf R}_i ) 
+\delta ({\bf r}-{\bf R}_i ) {\bf P}_i )  . \eqno{(2.2)}  $$
${\bf A}({\bf r})$ denotes the vector potential which is created by the 
atomic electron, and it can be written as 
$$ {\bf A}({\bf r})= \int { {\bf j}_L({\bf r}')
\over{|{\bf r}-{\bf r}' |}} d^3 r'  \eqno{(2.3)} $$
where ${\bf j}_L({\bf r})$ denotes the current density of the electron 
and is written as 
$$ {\bf j}_L({\bf r}') = (-e) {\bf \alpha} \delta ({\bf r}-{\bf r}')  . $$
In this case, the magnetic hyperfine splitting 
energy $W$ can be written as
$$W = <IJ:FF|H'|IJ:FF>= {1\over 2} [F(F+1)-I(I+1)-J(J+1)] a_I \eqno{(2.4)} $$ 
where $I$,$J$ and $F$ denote the  spin of the nucleus, the spin 
of the atomic electron and the total spin of the atomic system, 
respectively. $a_I$ is called the magnetic hyperfine structure (hfs) constant. 
Following ref.[3], we can write the expression for the $a_I$ as
$$ a_I =a_I^{(0)} (1+\epsilon ) \eqno{(2.5)} $$ 
where $a_I^{(0)}$ is the hfs constant for the point charge, and 
can be written as
$$ a_I^{(0)} = -{2ek\mu_N \over{IJ(J+1)}}
\mu  \int_0^\infty F^{(kJ)} G^{(kJ)} dr  \eqno{(2.6)} $$ 
where $F^{(kJ)}$ and $ G^{(kJ)}$ are the large and small components 
of the relativistic electron wave function for the $kJ$ state. 
$\mu_N$ is the nuclear magneton. 

$\epsilon$ is called the hfs anomaly and can be written as 
$$\epsilon =-{1\over{\mu}}<II|\sum_{i=1}^A N(R_i) 
\mu_i |II> -{1\over{\mu}}<II|\sum_{i=1}^A K(R_i) g_s^{(i)} 
\Sigma_i^{(1)} |II> \eqno{(2.7)} $$
where $ \Sigma_i^{(1)}$ is defined in eq.(1.1). 
$N(R)$ and $K(R)$ are written for the atomic electron as,
$$ N(R)=0.62 b^{(kJ)} \left( { R\over{R_0}} \right)^2 \eqno{(2.8)} $$
$$ K(R)=0.38 b^{(kJ)} \left( { R\over{R_0}} \right)^2 \eqno{(2.9)}$$
where $R_0$ is a nuclear radius and can be given as
$ R_0=r_0 A^{ 1\over 3} $ with $r_0=1.2 $ fm. 
On the other hand, $b^{(kJ)}$ is a constant 
which can be calculated in terms of relativistic electron 
wave functions and can be written as [3].  
$$ b^{(kJ)} = 0.23 k^2_0 R_0 \gamma (1-0.2 \gamma^2 ) \left[- (1+
{4R_0m_e c\over{3\gamma\hbar}}) \right] 
/ \int_0^{\infty} F^{(kJ)}G^{(kJ)} dr . \eqno{(2.10)} $$
$m_e$ denotes the electron mass, $k_0^2$ is a normalization constant and 
the factor in the square bracket is to be included for $p_{1\over 2}$ only, 
and $\gamma = Ze^2/\hbar c $. 

The isotope shift of the hfs anomalies of the two isotopes 
$\Delta_{12}$ is defined as 
$$ \Delta_{12} = {a_{I_1}g_2\over{a_{I_2}g_1}} -1   . \eqno{(2.11)} $$
Since the hfs anomaly $\epsilon$ is quite small, $\Delta_{12}$ 
becomes 
$$ \Delta_{12} \approx \epsilon_1 -\epsilon_2    . \eqno{(2.12)} $$

\vspace{3cm}

\item{\Large The hfs anomaly}

The hfs anomaly $\epsilon$ can be calculated if we know the 
nuclear wave function. However, it is often difficult 
and complicated 
to determine reliable nuclear wave functions. 

Here, we employ  simple-minded shell model wave functions with 
core polarizations taken into account. We take  
the following two approaches. The first one is to consider 
only the $\Delta \ell =0$ core polarization for the $ \Sigma_i^{(1)} $ 
operator. In this case, we can calculate the matrix element of 
the $ \Sigma_i^{(1)}$ without introducing any free parameters 
as discussed in ref.[3]. On the other hand, if we want to include 
 the  $\Delta \ell =2$ core polarization for the $ \Sigma_i^{(1)} $ 
operator, then we should use the nuclear wave function which can be 
obtained by truncating the shell model spaces. In the case of the Eu 
nuclei, the contribution of the  $\Delta \ell =2$ core polarization 
may be important since there are two orbits ( $2f_{7\over 2}$ 
and $1h_{9\over 2}$ ) nearby which generate a large effect on 
the matrix element of $[sY^{(2)}]_i^{(1)} $ operator. 

\newpage
\begin{enumerate}
\item{\large $\Delta \ell =0$ core polarization }

First, we consider the $\Delta \ell =0$ core polarization for 
the $ \Sigma_i^{(1)} $ operator [9-11]. 
In this case, we can express the effect of the core polarization 
on the $ \Sigma_i^{(1)} $ operator in terms of the core polarization 
of the magnetic moment. This is mainly because the core polarizations 
of $\sum g_s^{(i)} s_i $ and $\sum g_s^{(i)} \sqrt{2\pi} [sY^{(2)}]^{(1)}_i $ 
operators are related as
$$ \delta <\sum g_s^{(i)} \sqrt{2\pi} [sY^{(2)}]^{(1)}_i > = -{1\over 4} 
\delta <\sum g_s^{(i)} s_i >    .  $$
Therefore, we can write the expectation 
value of the $ \Sigma_i^{(1)} $ as 
$$ <II| \sum_{i=1}^A g_s^{(i)}  \Sigma_i^{(1)} |II> 
=\pm g_s^{(VN)}{3(I+{1\over 2})\over{4(I+1)}} + 
{3g_s^{(VN)}\over{4(g_s-g_{\ell})^{(VN)} }} (\mu-\mu_{sp}-
\delta \mu^{mes} ) \eqno{(3.1)} $$ 
for $I=\ell \pm {1\over 2} $ as discussed in ref. [3]. 
Here, $g_s^{(VN)}$ denotes the g-factor of the valence nucleon 
for the single particle state. $\mu_{sp}$ is 
the single particle value of the magnetic moment operator 
and can be written as 
$$ \mu_{s.p.}=(I-{1\over 2}) g_{\ell}^{(VN)} + {1\over 2}g_s^{(VN)} \quad  
 {\rm for} \quad I=\ell + {1\over 2}  
\eqno{(3.2a)} $$
$$ \mu_{s.p.}={I\over{I+1}} \left( (I+{3\over 2}) 
g_{\ell}^{(VN)} - {1\over 2}g_s^{(VN)} \right) \quad  
{\rm for} \quad I=\ell - {1\over 2}  . 
\eqno{(3.2b)} $$

$g_s^{(VN)}$ and $g_{\ell}^{(VN)}$ are taken to be 
 the free nucleon $g-$factors, 
$$ g_{\ell}^{(p)}=1.0, \qquad  g_s^{(p)}=5.5855 $$
$$ g_{\ell}^{(n)}=0, \qquad g_s^{(n)}=-3.8263   . $$
Also, $\delta \mu^{mes}$  is the effective magnetic moment 
arising from the meson exchange current [12] and can be approximated by 
$$ \delta \mu^{mes} \approx  0.1 \ell \tau_3   .  \eqno{(3.3)} $$
In the present calculation, however, we do not take into account 
the exchange current effects. This is because it is not very easy 
to calculate  the $\Delta \ell =2 $ core polarization effect 
together with the exchange current in a consistent way. In any case, 
the exchange current effects are not very large here. 

Therefore, we do not have any free parameters in the evaluation 
of the expectation value of the $ \Sigma_i^{(1)} $  for 
the $\Delta \ell =0$ core polarization case. 

\vspace{0.5cm}
\item{\large $\Delta \ell =2$ core polarization }

The contribution of the $\Delta \ell =2 $ core polarization 
to the operator $\sum g_s^{(i)} \Sigma_i^{(1)} $ 
depends very much on 
the nuclear configurations. For example, there is little chance 
for light nuclei that the $\Delta \ell =2$ core polarization 
becomes important. 
However, in the Eu nuclei, there may be a large contribution from the 
$\Delta \ell =2$ core polarization. This can be easily seen if we 
look at the neutron configurations. For example, in  the $^{147}Eu$ 
nucleus, the two neutrons outside the $N=82$ magic shell 
may have the following orbits, $2f_{7\over 2}$ 
and $1h_{9\over 2}$ nearby which have almost the same 
single particle energies. 
Therefore, there is a strong mixture between them 
due to the $ [sY^{(2)}]^{(1)}$ operator, and thus we have to consider 
these configurations  carefully.  

Since the Eu isotopes with even number of neutrons
 have the spin of $I={5\over 2}$, 
the proton state may be described by $2d_{5\over 2}^{-1}$. Also, 
the neutron number $N=82$ is a magic shell, and therefore the Eu 
nucleus with the mass number $A=63+82+n$ should have the $n$ 
neutrons outside the $N=82$ shell. Therefore, the wave function 
for the Eu nucleus with $n$ neutrons may be constructed by the 
following three states  
$$ |\Psi_0 : n>= \alpha_1 |1>+\alpha_2 |2>+ \alpha_3 |3> \eqno{(3.4)} $$
where
$$ |1>= |\pi (2d_{5\over 2})^{-1},
\nu (2f_{7\over 2})^{(n)}{(0^+)}: II > \eqno{(3.5a)}$$ 
$$ |2>= |\pi (2d_{5\over 2})^{-1},\nu \left( (2f_{7\over 2})^{(n-1)}
1h_{9\over 2} \right)_{(1^+)} :II > \eqno{(3.5b)} $$
$$ |3>= |\pi (2d_{5\over 2})^{-1},\nu \left( (2f_{7\over 2})^{(n-1)}
2f_{5\over 2} \right)_{(1^+)} :II >   .  \eqno{(3.5c)} $$
 $\alpha_1$, $\alpha_2$ and $\alpha_3$ should be determined 
by the nuclear residual interaction. Note that $I$ is here $5\over 2$. 

To determine the values of the $\alpha_i$, 
we employ the $\delta$-function force for the residual interaction 
for simplicity [11], 
$$V_{12}= -V_0 \delta ({\bf r}_1-{\bf r}_2)   .  \eqno{(3.6)} $$
Now, the problem is that we cannot use the perturbation theory here 
since the unperturbed energies of the  state $|1>$ 
and the state $|2>$ are degenerate. 
Therefore, we should diagonalize the hamiltonian with the residual 
interaction. Denoting the unperturbed energies for the states $|i>$ 
as $ E_i$, we obtain the matrix equations which determine the 
values of $\alpha_i$. 
$$ (E_1 +V_{11}-E) \alpha_1 +V_{12}\alpha_2+V_{13}\alpha_3 =0 \eqno{(3.7a)}$$
$$ V_{21}\alpha_1+ (E_2 +V_{22}-E) \alpha_2 +V_{23}\alpha_3 =0 \eqno{(3.7b)} $$
$$ V_{31}\alpha_1+V_{32}\alpha_2+(E_3 +V_{33}-E) \alpha_3  =0 . \eqno{(3.7c)}$$
Here, we can take $E_1 =E_2$ to a good approximation. Also, we can 
neglect the interference term between the $|2>$ and $|3>$ states. 


Before going to determine the values of the parameters $V_0$ and $E_1$, 
we discuss the core polarizations which contribute to the magnetic 
moment. Since the N=82 is the magic shell, the neutrons of 
the $^{145}Eu$ nucleus are all filled. In  this case, there is no 
contribution from the $\Delta \ell =2$ core polarization. Instead, 
the $\Delta \ell =0$ core polarization comes from the neutrons 
configuration mixing of 
$$ |\pi (2d_{5\over 2})^{-1},\nu \left( (1h_{11\over 2})^{-1}
1h_{9\over 2} \right)_{(1^+)} :II >  $$ 
 state, and the proton configuration mixing of 
$$ |\pi (2d_{5\over 2})^{-2}, (2d_{5\over 2}) :II > $$
state. We take these effects perturbatively. 

For the heavier Eu isotopes, we assume that this part of the  
$\Delta \ell =0$ core polarization behaves just in the same way as 
the $^{145}Eu$ nucleus. Therefore, the nuclear state of eq.(3.4) 
should reproduce the magnetic moment which is the observed 
magnetic moment plus the contribution from 
the $\Delta \ell =0$ core polarization. This can be easily obtained 
since the $^{145}Eu$ has only the $\Delta \ell =0$ core polarization, 
which is $\delta \mu_{CP} = -0.8$ n.m. 

However, this is possible only up to the $A=151$ nucleus, and  
 there is a big change of the observed magnetic moment from $A=151$ 
to $A=153$. The magnetic moment of $A=153$ nucleus is
 smaller than $A=151$ by a factor 2. This means that  
  we cannot reproduce the magnetic moment of the nuclei 
  heavier than the  $A=153$ by the 
 simple-minded core polarization. 

Therefore, we should consider some kind of collective state (deformed state) 
into the original state $ |\Psi_0 : n>$. We call this new state 
$|MAS>$ state (magnetic absorbing state) since the $|MAS>$ state 
is assumed to have  
a vanishing magnetic moment. That is, 
$$ <MAS | \sum_{i=1}^A \mu_i | MAS > =0   . \eqno{(3.8a)} $$
At the same time, we assume that the expectation value of the 
operator $\sum g_s^{(i)} \Sigma_i^{(1)} $ with the MAS state vanishes,
$$ <MAS | \sum_{i=1}^A g_s^{(i)} \Sigma_i^{(1)} | MAS> = 0 .   \eqno{(3.8b)} $$
In this case, 
we can construct the wave functions for the $^{153}Eu$ nuclei. 
We write 
$$ |\Psi :MAS >= \sqrt{1-\alpha_4^2} |\Psi_0:n=8> + \alpha_4 |MAS>   .
 \eqno{(3.9)} $$
 
Here, the value of the parameter $\alpha_4$ can be determined such that 
the observed magnetic moment can be reproduced for the $^{153}Eu$ nuclei. 

Now, concerning the configurations 
for the A=155 and heavier nuclei, the $2f_{7\over 2}$ states  are filled. 
Therefore, we take the following configuration 
$$ |\widetilde{ \Psi_0} : n>= \alpha_1 |\widetilde{1}> 
+\alpha_2 |\widetilde{2}>+ \alpha_3 |\widetilde{3}> \eqno{(3.10)} $$
where
$$ |\widetilde{1}>= |\pi (2d_{5\over 2})^{-1},
\nu (1h_{9\over 2})^{(n-8)}{(0^+)}: II > \eqno{(3.11a)}$$ 
$$ |\widetilde{2}>= |\pi (2d_{5\over 2})^{-1}, 
\nu \left( (1h_{9\over 2})^{(n-7)}
(2f_{7\over 2})^{-1} \right)_{(1^+)} :II > \eqno{(3.11b)} $$
$$ |\widetilde{3}>= |\pi (2d_{5\over 2})^{-1}, 
\nu \left( (2f_{7\over 2})^{-1}
2f_{5\over 2} \right)_{(1^+)} :II >   .  \eqno{(3.11c)} $$ 
Here, the value of the $\alpha_3$ is fixed to the one determined 
for A=153. 
Further, the wave function is taken to be  
$$ |\Psi :MAS >= \sqrt{1-\alpha_4^2} |\widetilde{ \Psi_0} : n> 
+ \alpha_4 |MAS>   . \eqno{(3.12)} $$
These are the shell model wave functions which we use for our calculations. 
However, they are obviously too simple-minded wave functions, but 
we believe that we can get some idea as to what are the contributions 
from the $\Delta \ell =2$ core polarization as well as from the sharp 
change of the magnetic moment to the hfs constant in the Eu isotopes.


\vspace{0.5cm}

\item{\large Moskowitz-Lombardi empirical formula }

Moskowitz and Lombardi proposed an empirical formula in order 
to explain the isotope shifts of the hfs anomaly in Hg isotopes [5]. 
They simply take the shape of the hfs anomaly to be  
$$ \epsilon = {\alpha\over{\mu}} $$
where $\alpha$ is a constant. When they take $\alpha$ to be $\pm 0.01 $ n.m. 
for $j=\ell \pm {1\over 2}$ neutron orbits, they obtain a remarkably 
good description of the isotope shifts $\Delta_{12}$ 
for the Hg nuclei. 

Here, we want to apply this formula to the Eu isotopes. In this case, we 
should make a correction due to the atomic orbit. It should reflect 
in the value of the $\alpha $ since the hfs anomaly depends on $b^{(kJ)}$. 
Therefore, we make this atomic correction as 
$$ C= {b^{(kJ)}(Eu)\over{b^{(kJ)}(Hg)}}  \approx 0.5   . $$ 
Therefore, we take the $\alpha$ to be $\mp 0.005 $ n.m. for 
$j= \ell \pm {1\over 2} $ proton orbits.

\end{enumerate}

\vspace{3cm}
\item{\large Numerical Results}

Once we know the values of $\alpha_i$ in eqs.(3.5) and (3.11), 
then we can calculate 
the expectation values of the magnetic moment as well as  the 
$ \Sigma_i^{(1)} $ operator. 

Before going to the numerical calculations, we make the valence 
nucleon approximation to the expectation values of $< ({R_i\over{R_0}})^2 >$. 
Since the dominant contributions to the magnetic moment as well as to the 
$ \Sigma_i^{(1)} $ operator come from the valence nucleons, it is always 
a good approximation to factorize the $< ({R_i\over{R_0}})^2 >$ and 
$\mu_i$ or $ \Sigma_i^{(1)} $. Therefore, eq.(2.7) can be written as 
$$\epsilon =-0.62 b^{(kJ)} < \left( {R_i\over{R_0}} \right)^2>_{VN}  
-0.38 b^{(kJ)} < \left( {R_i\over{R_0}} \right)^2>_{VN}
{1\over{\mu}}<II|\sum_{i=1}^A  g_s^{(i)} 
\Sigma_i^{(1)} |II> \eqno{(4.1)} $$
where $ < ({R_i\over{R_0}})^2>_{VN}  $ is the expectation value 
with the valence nucleons. 

Now, we can calculate the hfs anomaly with the shell model 
wave functions. First, we consider the $\Delta \ell =0 $ core polarizations. 
  In this case, we do not have any free parameters 
since all the terms in eq.(3.1) are known.  
In Table 1, we show our calculated results of 
 the hfs anomaly with the $\Delta \ell =0 $ 
 core polarization for the Eu nuclei. 
Also, the calculated results of the isotope shifts $\Delta_{145,A}$ are 
shown. 

Next, we consider the $\Delta \ell =2 $ core polarizations. 
In this case, we should first determine the values of $\alpha_i$. 
Basically we change the value of the $E_3$ so that we can obtain 
the observed magnetic moment. 

In Table 2, we show our calculated results of the hfs anomaly 
including the $\Delta \ell = 2$ core polarizations. Also, the isotope shifts 
of the hfs anomaly are shown. Here, as mentioned before, 
we have not considered 
the exchange current effect on the hfs anomaly. Therefore, there is 
no difference in the hfs anomaly for the $^{145}Eu$ between 
Table 1 and Table 2 since $^{145}Eu$ does not have any effects 
due to the $\Delta \ell =2 $ core polarizations. On the other hand, 
the difference in the hfs anomaly  between Table 1 and Table 2 
for the heavier $Eu$ isotopes is partly due to the $\Delta \ell =2$ 
core polarization effect and partly due to the MAS state. 
In particular, the large difference between the two calculations 
for the Eu nuclei heavier than the A=153 comes from the MAS state. 
The $\Delta \ell =2$ core polarization 
contribution to the hfs anomaly for the $^{147}Eu$ is about 
$\epsilon^{\Delta \ell =2} \approx 0.01 \quad \% $.  

These calculations should be compared to the prediction 
by Moskowitz-Lombardi empirical rule of eq.(1.2). Note that the 
large difference of the hfs anomaly $\epsilon$ in magnitude 
between the present calculations and the M-L formula is mainly 
due to the first term in eq.(4.1). However, the first term in eq.(4.1) 
contributes very little to the isotope shifts $\Delta_{12}$ since 
it is practically constant. 

In order to see more clearly the difference between the models, we plot 
in fig. 1 the isotope shifts of the hfs anomaly 
as the function of the nuclear mass number $A$. 

First, we want to present the results of the 
Mercury isotopes so that we can obtain some idea how these models can 
describe the data. 
In fig. 1a, we show the predictions 
of the $\Delta_{199,A}$ for the Hg nuclei by the different models. 
There, the difference between the model calculations is not very large. 
At least, there is a qualitative agreement in the behavior, although 
the M-L formula (dashed line) 
 can describe the data better than the shell model
calculations. The FULL (solid line) in fig. 1a indicates that the calculation 
includes both the $\Delta \ell = 0$ and $\Delta \ell =2 $ core polarizations. 
Here, however, the $\Delta \ell =2 $ core polarization is not very large [3].  

In fig.1b, we show the calculated results for the Eu nuclei. 
In this case, there is a big difference between the
model calculations of the isotope shifts $\Delta_{145,A}$. 
The Moskowitz-Lombardi empirical formula (dashed line) 
 predicts a big change 
in the isotope shifts $\Delta_{145,A}$ at the $A=153$ nucleus. 
On the other hand, our theoretical estimations show a very different 
behavior  of the isotope shift $\Delta_{145,A}$. 
In particular, the predictions with the $\Delta \ell =0$ core 
polarizations (dotted line) give a completely opposite 
behavior to the M-L result. 
The sign of the isotope shifts $ \Delta_{145,A}$ are negative. 
Also, 
the FULL shell model calculations (solid line) show a quite smooth 
transition at A=153. This is mainly due to the introduction 
of the MAS state. 

Therefore, it should be extremely interesting to check  
experimentally which of the models can be reasonable. 
Further, it would be very interesting to understand whether 
there is any universality in the M-L formula. 

\newpage
\item{\large Conclusions} 

We have presented the numerical calculations of the isotope 
shifts of the magnetic hfs anomaly for the Eu nuclei. 
Here, we have considered the $\Delta \ell =0 $ as well as the 
$\Delta \ell =2 $ core polarizations to the hfs operators. 
It turns out that the calculations with both of the core polarizations 
do not show any sharp transition at $A=153$ where the observed magnetic 
moment show a big change. This is because the hfs operators and the 
magnetic moment operators behave similarly by the core polarizations, 
and therefore the effects of the core polarizations cancel with each 
other. 

On the other hand, there is a very nice empirical formula by 
Moskowitz-Lombardi, which perfectly describes the isotope shifts 
of the hfs anomaly in Hg nuclei. Here, we also employ the M-L formula 
how it predicts the isotope shifts of the hfs anomaly 
in the Eu nuclei. It turns out that the M-L formula predicts a  
transition at $A=153$ since this is essentially proportional 
to the inverse of the magnetic moment. 

Therefore, it would be extremely nice to learn 
which of the pictures the nature prefers. Also, it is quite interesting 
to understand if there is any universality in this simple M-L rule.

\end{enumerate}

\vspace{3cm}

Acknowledgment: We thank  I. Katayama and G. Werth for discussions. 
This work is supported in part by Japanese-German Cooperative Science 
Promotion Program.

\newpage

\underline{\large References}
\vspace{0.5cm}

1. A. Bohr and V.F. Weisskopf, Phys. Rev. {\bf 77}, 94 (1950) 

2. J. Johnson and R.A. Sorensen, Phys. Rev. {\bf C2}, 102 (1970)

3. T. Fujita and A. Arima, Nucl. Phys. {\bf A254}, 513 (1975)

4. S. Buttgenbach, Hyperfine Structure in 4d and 5d-Shell Atoms, 

\ \ \ \ \  Springer Tracts in Modern Physics {\bf 96} (1982)

5. P.A. Moskowitz and M. Lombardi, Phys. Lett. {\bf 46B}, 334 (1973)

6. G. Werth, S. Trapp, G. Bollen, S. Schwarz and A.Kohl, CERN/ISC 96-9  

7. M. Wada, K. Okada, H. Wang, K. Enders, F. Kurth, T. Nakamura, S. Fujitaka, 

\quad  J. Tanaka, H. Kawakami, S. Ohtani and I. Katayama, 
to appear in Nucl. Phys. {\bf A}

8. K. Enders, O. Becker, L. Brand, J. Dembczynski, G. Marz, G. Revalde, 

\quad P.B. Rao and G. Werth, Phys. Rev. {\bf A52}, 4434 (1995)

9. A. Arima and H. Horie, Prog. Theor. Phys. {\bf 12}, 623 (1954)

10. R.J. Blin-Stoyle and M.A. Perks, Proc. Phys. Soc. {\bf A67}, 885 (1954)

11. H. Noya, A. Arima and H. Horie, Prog. Theor. Phys. Suppl. {\bf 8}, 33 (1958)

12. S. Nagamiya and T. Yamazaki, Phys. Rev. {\bf C4}, 1961 (1971)

13. R.L. King, C.H. Liu, G. Fulop and H.H. Stroke, 
Phys. Lett. {\bf B31}, 567 (1970)
 
14. P.A. Moskowitz, C.H. Liu, G. Fulop and H.H Stroke, 
Phys. Rev. {\bf C4}, 620 (1971)

\newpage
\begin{center}
\underline{Table 1} \\
$\displaystyle{ \rm The \  hfs \  anomaly \  for \  Eu \ isotopes 
  } \qquad (\Delta \ell = 0)  $ \\
\ \ \ \\
\begin{tabular}{|c|c|c|c|c|}
\hline
A & $\mu$ & $b^{(1s_{1\over 2})} \quad (\%) $  & $ - \epsilon \quad (\%) $ 
& $  \Delta_{145,A} \quad (\%) $ \\
\hline
\hline
145 & 3.993 & 1.58 & 1.067 & 0 \\
\hline
147 & 3.724 & 1.58 & 1.056 & $-$0.011 \\
\hline
149 & 3.565 & 1.59 & 1.053 & $-$0.014 \\
\hline
151 & 3.472 & 1.59 & 1.046 & $-$0.021  \\
\hline
153 & 1.533 & 1.60 & 0.919  & $-$ 0.148 \\
\hline 
155 & 1.56 & 1.60 & 0.919 & $-$0.148 \\
\hline 
157 & 1.5 & 1.61 & 0.912  & $-$0.155 \\
\hline
159 & 1.38 & 1.62 & 0.893 & $-$0.174 \\
\hline
\end{tabular}

\vspace{1cm}
\begin{minipage}{13cm}
We plot the hfs anomaly and the isotope shifts 
for the Eu isotopes with the $\Delta \ell =0$ 
core polarization. In addition, we show the magnetic moment and 
the electron wave function coefficient $b^{(kJ)}$ for $1s_{1\over 2}$. 

\end{minipage}


%
%

\vspace{3cm}
\underline{Table 2} \\
$\displaystyle{ \rm The \  hfs \  anomaly \  for \  Eu \ isotopes 
  } \qquad (\Delta \ell = 0) \ \  + \ \   (\Delta \ell = 2)  $ \\
\ \ \ \\
\begin{tabular}{|c|c|c|c|c||c|c|}
\hline
A & $\mu$ & $b^{(1s_{1\over 2})} \quad (\%) $  & $ - \epsilon \quad (\%) $ 
& $  \Delta_{145,A} \quad (\%) $ & $-\epsilon^{ML} \quad (\%)$ & 
$\Delta_{145,A}^{ML} \quad (\%) $ \\
\hline
\hline
145 & 3.993 & 1.58 & 1.067 & 0 & 0.125 & 0 \\
\hline
147 & 3.724 & 1.58 & 1.044 & $-$0.023 & 0.1349 & 0.009 \\
\hline
149 & 3.565 & 1.59 & 1.042 & $-$0.025 & 0.140 & 0.015 \\
\hline
151 & 3.472 & 1.59 & 1.036 & $-$0.031 & 0.144 & 0.019  \\
\hline
153 & 1.533 & 1.60 & 1.039  & $-$0.028 & 0.326 & 0.201 \\
\hline 
155 & 1.56 & 1.60 & 1.035 & $-$0.032 & 0.32 & 0.20 \\
\hline 
157 & 1.5 & 1.61 & 1.019  & $-$0.048 & 0.33 & 0.21 \\
\hline
159 & 1.38 & 1.62 & 1.019 & $-$0.048 & 0.36 & 0.24 \\
\hline
\end{tabular}

\vspace{1cm}
\begin{minipage}{13cm}
We plot the hfs anomaly and the isotope shifts 
for the Eu isotopes with the $\Delta \ell =0$ and $\Delta \ell =2 $  
core polarizations. Also, we show the predictions 
by the M-L empirical formula with the universal coupling of 
$\alpha = -0.005 $ n.m. 

\end{minipage}

\end{center}

\newpage

\underline{\large  Figure Captions : }

\begin{list}{}{}
\item[Fig.1a] : 

The isotope shifts $\Delta_{199,A}$ of the hfs anomaly 
for the Hg nuclei are shown as the function of the mass number. 
The solid line (FULL) denotes the shell model calculation including 
both the $\Delta \ell =0$ and $\Delta \ell =2$ core polarizations. 
The dashed line (M-L) is the calculation by the M-L empirical formula 
( $\epsilon^{ML}$ and $\Delta^{ML}_{A,145}$ ). 
The black circles denote the experiment [13,14].  

\vspace{0.5cm}

\item[Fig.1b] : 

The isotope shifts $\Delta_{145,A}$ of the hfs anomaly 
for the Eu nuclei are shown
The same as fig.1a for the solid and dashed lines. 
The dotted line denotes the calculation only with 
the $\Delta \ell =0$ core polarization. 

\end{list} 

\end{document}